\documentclass{iopart}
\usepackage{graphicx}

\newcommand{\dbar}{{\mathchar'26\mkern-12mu d}}

\begin{document}
\begin{flushright}
YITP-08-32
\end{flushright}

\title{Chiral Symmetry and Heavy-Ion Collisions}
\author{Kenji Fukushima}
\address{Yukawa Institute for Theoretical Physics,\\
         Kyoto University, Kyoto 606-8502, Japan}
\begin{abstract}
 I revisit the phase structure of hot and dense matter out of quarks
 and gluons with some historical consideration on the color
 deconfinement and chiral phase transitions.  My goal is to make clear
 which part of the QCD phase diagram is under theoretical control and
 which part is not.  I demonstrate that an uncommon but logically
 possible scenario other than the standard phase diagram cannot be
 ruled out.  My emphasis is on the concern that one should correctly
 understand what kind of phenomenon occurs associated with the phase
 boundary line on the diagram.  It is not quite obvious, in
 particular, where chiral symmetry restoration plays a
 phenomenological role in the temperature and baryon density plane
 except at the QCD (chiral) critical point.
\end{abstract}


\section{How is the QCD phase diagram built?}

  The quest for a new state of matter out of quarks and gluons is
becoming a matured subject in physics founded on Quantum
Chromodynamics (QCD).  We are opening an epoch of exact science for a
quark-gluon plasma.  There have already come convincing experimental
data from RHIC at the Brookhaven National Laboratory and more will be
forthcoming from LHC at CERN in a couple of years.  So far, the
intrinsic properties of transient matter are mainly concerned from the
experimental point of view, but an enough deal of compilation of
experimental outputs would enable us to access the detailed structure
of the QCD phase transitions directly.

  In fact, it is the conjectured QCD phase diagram that has been a
guiding principle leading to feasible research projects to create hot
and dense QCD matter in laboratory.  Since the standard shape of the
QCD phase diagram is widely prevailing as a common knowledge, it is
more or less difficult for us to doubt such an almost unanimous
assent.  If one tries to figure it out somehow from theory, however,
one should realize that our current knowledge does not suffice to
guarantee that the standard scenario is a unique possibility.  To
understand this it would be instructive to start with a historical
overview of how the QCD phase diagram is built.

  There is a consensus in the community that the very first QCD phase
diagram with the temperature $T$ and the baryon density $n_{\rm B}$
axes appeared in a paper by Cabibbo and Parisi in
1975~\cite{Cabibbo:1975ig}.  They interpreted the exponentially
increasing spectrum by Hagedorn (that is, the Hagedorn spectrum) as
``deconfinement'' and insisted that quarks are confined in the
low-$T$ and low-$n_{\rm B}$ region and deconfined otherwise.  The
``quarter sector'' shape of the phase boundary line, which is quite
familiar to us today, first revealed itself then.  I would like to
remind that chiral symmetry plays no role at all in this context.  Of
course, the theory of chiral symmetry restoration was developing
almost parallelly but separately from the color deconfinement theory.

\begin{figure}
 \begin{center}
 \includegraphics[width=12cm]{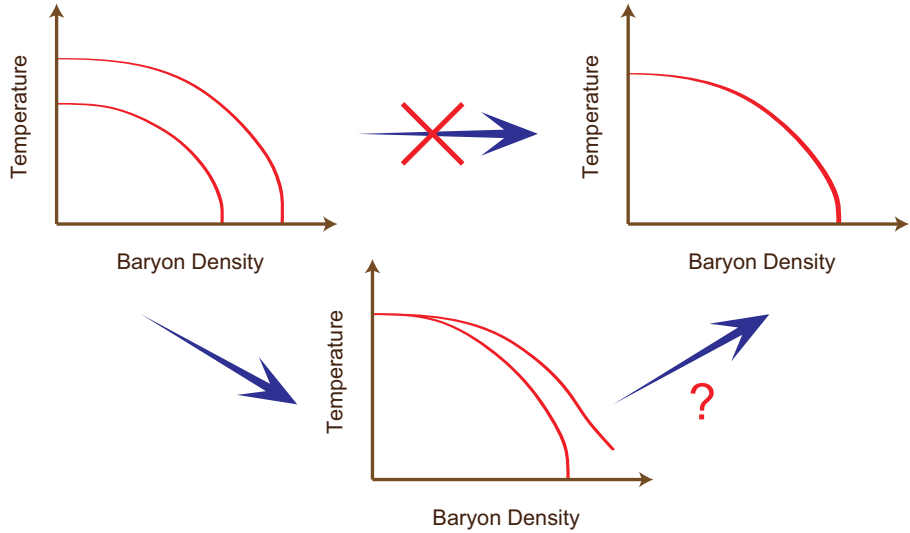}
 \end{center}
\caption{Schematic evolution of the possible phase diagrams in the
  temperature $T$ and baryon density $n_{\rm B}$ (or chemical
  potential $\mu$) plane.  Two red curves represent the deconfinement
  and chiral restoration phase boundaries.  For simplicity the
  mixed-phase region associated with the first-order transition is not
  considered or the horizontal axis should be not $n_{\rm B}$ but
  $\mu$, which was not taken quite seriously in the 80's.}
\label{fig:conventional}
\end{figure}

  Then, a dramatic leap, in my opinion, was made in 1983 into a
prototype of what we now know as the QCD phase diagram.  Let me select
out two historic figures depicted both by Baym.  One is found in his
contribution to the Bielefeld proceedings published in
1982~\cite{Baym:1982sn}.  It might look bizarre to draw two distinct
boundary lines on the phase diagram like shown in the upper-left of
Fig.~\ref{fig:conventional}, which is rather exotic to us but was a
serious candidate in 1982.  Each boundary line represents either
deconfinement or chiral symmetry restoration.  Another important and
more famous figure is put in the NSAC Long Range Plan for Nuclear
Science in 1983 which is readily available also from a nice review
article on the RHIC history~\cite{Baym:2001in}.  This figure had
undergone an important evolution from a double-boundary phase diagram
to a single-boundary shape like sketched in the upper-right of
Fig.~\ref{fig:conventional}.

  What caused such a qualitative difference between 1982 and 1983?
The answer lies in a situational change in the lattice Monte-Carlo
simulation by the Illinois group published in
1983~\cite{Kogut:1982rt}.  It was claimed that deconfinement and
chiral symmetry restoration take place at almost the same temperature
in both SU(2) and SU(3) gauge theories.  The locking of two
transitions or crossovers has been confirmed by more accurate
simulations, which makes most people in the community believe in the
single-boundary phase diagram.  [Whether the locking is complete or
approximate is still under dispute~\cite{karsch,fodor}, but in any
case two critical temperatures are close to each other.]

  This deduction, however, skips one logical step, I would say.  Our
knowledge from the Monte-Carlo simulation is limited to the region
where the baryon density is much smaller than the temperature.
Therefore, the single-boundary diagram is too severely constrained
than what we can conclude from the lattice results.  The logical
deduction from the upper-left of Fig.~\ref{fig:conventional} and the
coincidence of two critical temperatures at zero density leads to a
phase diagram as given in the lower-middle of
Fig.~\ref{fig:conventional}.  I would dare not deny the conventional
phase diagram, however, by this argument;  in
Fig.~\ref{fig:conventional} the upper-right is still a possible
candidate as a special reduction from the lower-middle.  My point is
that we have not discovered any evidence for such a reduction.

  We shall assume a double-boundary phase diagram for the moment.
Then, we would hit on the following question;  which line is which of
deconfinement and chiral symmetry restoration?  Unfortunately no
theoretical argument can tell it in a rigorous way.  Some
discussions~\cite{Casher:1979vw,tho80} imply that quark confinement
should be accompanied by the spontaneous breaking of chiral symmetry
in the vacuum, but these theoretical arguments do not hold in a
medium.  I will present some model analyses on what will transpire
later.

  Here let us turn to another big stream toward the QCD phase
diagram.  That is, a pile of achievements based on the low-energy
chiral effective model.  The most prominent finding from model
analyses is, as I will explain below, the critical end-point that is
sometimes referred to as the QCD critical point.  Since I cannot
enumerate all the chiral effective models such as the linear and
non-linear sigma models, the Nambu--Jona-Lasinio (NJL) model, the
chiral random matrix model, the instanton liquid model, and so on, I
will focus only on the NJL model in this article.  This is because the
QCD critical point was first pointed out in a work based on the NJL
model~\cite{Asakawa:1989bq} and also because my discussions will be
followed later by state-of-the-art results in a closely related model
with the Polyakov loop incorporated.

  One of the widely-known early works by means of the NJL model is
that done by Hatsuda and Kunihiro~\cite{Hatsuda:1985eb}.  They drew a
line on the $T$ and $\mu$ plane in 1985 that separates the chiral
symmetric and broken phases with an explicit statement that they
neglected any effect of confinement.  Note that the NJL model at
finite $T$ is premised on deconfined matter in terms of quarks.  In
1986 Asakawa and Yazaki got aware of the first-order phase transition
at high density.  That means that the chiral phase transition is
continuous in the small-$\mu$ region and discontinuous in the
large-$\mu$ region.  There must be a resultant terminal point where
the phase transition ceases to be of first-order.  This end-point of
the first-order boundary is commonly called the chiral critical
end-point (CEP) or simply the QCD critical point.  I think that the
former makes more sense though the latter seems more prevailing.  I
would emphasize that the so-called QCD critical point has an origin in
chiral dynamics alone and no direct link to confinement and
deconfinement.

  In fact, it is extremely important to keep in mind the fact that the
QCD critical point is associated with chiral symmetry.  For instance,
our limited knowledge cannot exclude the following scenario however
unlikely (and undesirable) it sounds.  The density effect, as I
mentioned, could make the deconfinement phase transition take place in
a totally different way from chiral restoration.  If the QCD critical
point sitting on the chiral transition line is located far away from
deconfinement which predominantly determines the equation of state, it
should be hardly conceivable to expect any significant influence on
the hydrodynamic path even near the QCD critical point.  This is still
logically plausible.

  We have looked over theoretical understanding and remaining
possibilities of the QCD phase diagram on a general footing.  We shall
now proceed into concrete studies using the effective model in the
subsequent sections.


\section{Is it deconfinement or chiral restoration?}

  In general there should be a physical quantity which behaves
differently from one phase to the other to define the phase
transition.  It is the chiral condensate $\langle\bar{q}q\rangle$ in
the case of the chiral symmetry breaking and the Polyakov loop $\ell$
serves as an order parameter for deconfinement.  That is,
\begin{eqnarray*}
 &&\mbox{Chiral Symmetric Phase} \;\to\;
  \langle\bar{q}q\rangle = 0 \,,\\
 &&\mbox{Chiral Broken Phase} \;\to\;
  \langle\bar{q}q\rangle \neq 0 \,,\\
 &&\mbox{Quark Confinement Phase} \;\to\;
  \ell = 0 \,,\\
 &&\mbox{Quark Deconfinement Phase} \;\to\;
  \ell \neq 0 \,.
\end{eqnarray*}
Theoretically speaking, however, $\langle\bar{q}q\rangle$ could be
exactly zero only when chiral symmetry is exact, i.e., the quark mass
$m_q$ is strictly zero.  As for deconfinement, similarly, $\ell$ could
be zero only when $m_q$ is infinitely large (that is, center symmetry
is exact in the quenched limit).  Therefore none of the above can
define an exact critical point in the real world with finite quark
masses.  Nevertheless, it is still sensible to locate the
pseudo-critical point by the peak position of the susceptibility.
Remember that the susceptibility with respect to order parameter
diverges at the second-order critical point, and thus, it may well
have a maximum even in the case of smooth crossover.

  Because we have two independent order parameters, it would be rather
natural to anticipate two distinct pseudo-critical temperatures.  The
fact is, however, that the lattice QCD simulations result in the
coincidence of two peak positions of the chiral and Polyakov loop
susceptibility.  It is quite hard to explain this coincidence unless
two order parameters are mixed up to form one order
parameter~\cite{Hatta:2003ga}.  I would stress that the mixing
argument is not adequate to exclude two peaks in respective
susceptibility~\cite{Digal:2000ar,Mocsy:2003qw}.

  Here I will introduce one simple idea that turns out to be quite
successful to reproduce simultaneous crossover in terms of
$\langle\bar{q}q\rangle$ and $\ell$ as a function of $T$.  A textbook
knowledge on statistical mechanics reads the partition function for
free quarks as $Z_f=2\int\dbar p\{\log[1+e^{-(E-\mu)/T}]^3
+\log[1+e^{-(E+\mu)/T}]^3\}$ with 2 and 3 being spin and color
factors, which is modified by the presence of the Polyakov loop into a
form of~\cite{Fukushima:2003fw,Ratti:2005jh,Fukushima:2008wg}
\begin{eqnarray*}
 Z_f &=& 2\int\dbar p\;\Bigl\{\log\bigl[ 1+ 3\,\ell\, e^{-(E-\mu)/T}
  + \,3\,\bar{\ell} e^{-2(E-\mu)/T} + e^{-3(E-\mu)/T} \bigr] \\
 &&+ \log\bigl[ 1+ 3\,\bar{\ell}\, e^{-(E+\mu)/T}
  + 3\,\ell\, e^{-2(E+\mu)/T} + e^{-3(E+\mu)/T} \bigr]\Bigr\} \,,
\end{eqnarray*}
where $e^{-(E-\mu)/T}$ represents one-quark excitation which couples
$\ell$ and $e^{-2(E-\mu)/T}$ represents two-quark excitation which has
the same color structure as one-anti-quark excitation and thus couples
$\bar{\ell}$.  The last term, $e^{-3(E-\mu)/T}$, represents the
colorless baryon-like excitation.  The second line corresponds to the
anti-quark contribution.  Because the energy dispersion relation $E$
depends on the chiral condensate $\langle\bar{q}q\rangle$ through the
constituent quark mass, the coupling between $\ell$ and
$\langle\bar{q}q\rangle$ weakens when chiral symmetry is heavily
broken.  Conversely, as long as $\ell$ is vanishingly small, only the
baryon-like excitation is allowed and the chiral phase transition is
delayed accordingly.  These dynamical entanglement leads to
simultaneous crossover of deconfinement and chiral restoration.

  We see immediately that the nature of confinement is unchanged for
any $\mu$ as long as $T$ is sufficiently small.  This is because only
the term $e^{-3(E-\mu)/T}$ becomes dominant at $\mu>E$ and $T\simeq0$,
and all the coupling terms to $\ell$ drop off.  The physical
interpretation for this is transparent in fact.  Let us imagine that
we have free quark matter at a certain value of $\mu$.  Then, three
degenerate quarks with different colors should occupy each level in
the phase space up to the Fermi surface and they form a color singlet
in momentum space.  It does not have to be a spatially compact object
like a nucleon.  Hence, quark deconfinement in configuration space
would make no further qualitative change to degenerate quark matter if
seen in momentum space.

  As a result of persisting confinement at low $\mu$ the upper and
lower lines in the lower-middle of Fig.~\ref{fig:conventional} should
correspond to deconfinement and chiral restoration, respectively,
according to this intuitive argument.  That is actually the case in
the model study~\cite{Fukushima:2008wg}.


\section{What can theoretical models tell?}

\begin{figure}
 \includegraphics[width=6cm]{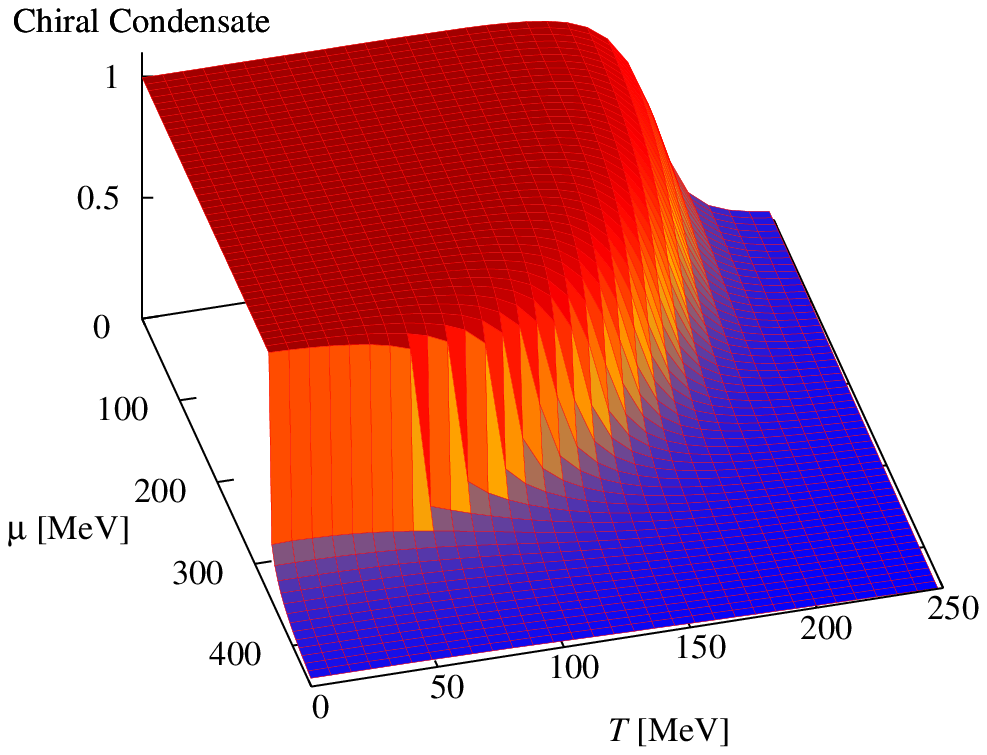}
 \hspace{5mm}
 \includegraphics[width=6cm]{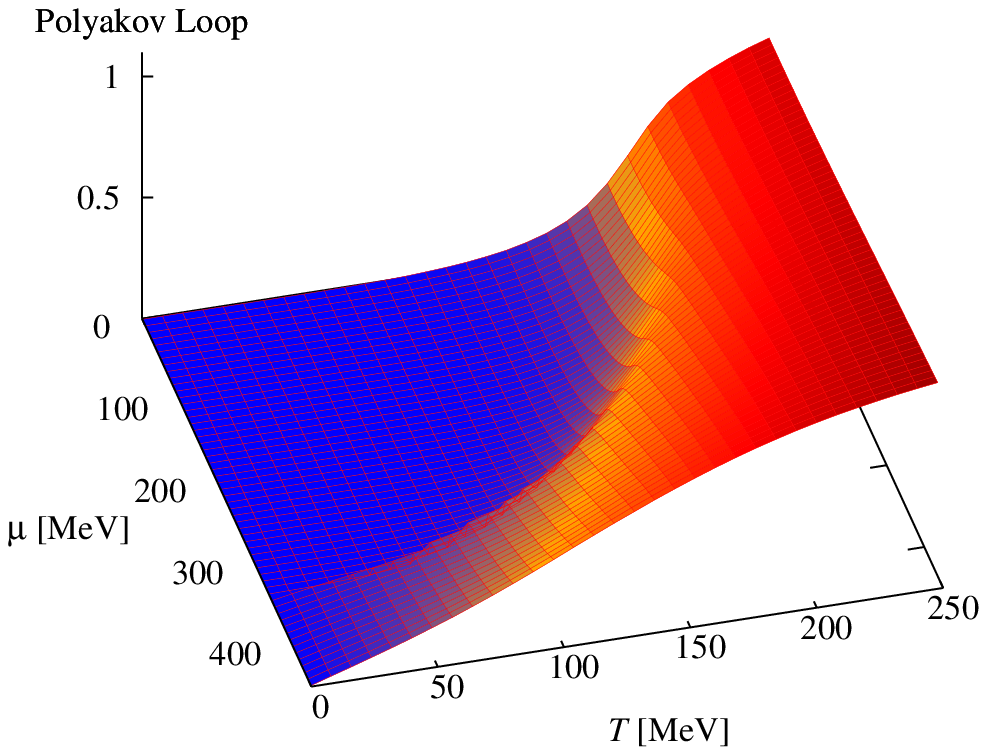}
 \caption{Behavior of order parameters in the $T$-$\mu$ plane in the
 PNJL model taken from Ref.~\cite{Fukushima:2008wg}.  Left) Chiral
 condensate for light (\textit{up} and \textit{down}) quarks
 normalized by the value at $T=\mu=0$.  Right) Polyakov loop $\ell$.}
 \label{fig:order}
\end{figure}

  In the rest of this article I will address the QCD phase diagram
inferred from the recent analyses in the three-flavor NJL model with
the Polyakov loop augmented (PNJL model).  We see the behavior of the
light-quark chiral condensate $\langle\bar{u}u\rangle$ (normalized by
the vacuum value) and the Polyakov loop $\ell$ in the $T$-$\mu$ plane
in Fig.~\ref{fig:order}.  Here it should be reiterated that the model
parameters are fixed to reproduce the vacuum property and the
simultaneous crossover at $T\neq0$ and $\mu=0$.  All the data at
$\mu\neq0$ are thus the model prediction that is to be compared to the
lattice QCD simulation in the future when the sign problem will be
solved.  I remark that color superconductivity is not taken into
account in the results displayed in Fig.~\ref{fig:order}.

  It is clear that the chiral condensate and the Polyakov loop are
closely related to each other.  We might consider at a first glance
that the upper-right of Fig.~\ref{fig:conventional}, that is, the
standard single-boundary phase diagram might be concluded from these
two plots in Fig.~\ref{fig:order}.  That is, however, not a right
interpretation.

  The chiral condensate has a discontinuous jump in the low-$T$ and
high-$\mu$ region.  We can perceive a tiny jump in the Polyakov loop
as well.  This is because of mixing between two order parameters.  As
we understood, however, the Polyakov loop cannot go large at
$T\simeq0$ regardless of the first-order phase transition.  That means
that the phase transition signifies a change from a chiral broken
confined phase to a chiral symmetric confined phase, if we naively
regard $\ell=0$ as confinement.  Accordingly the phase structure
stemming from this kind of model approaches has no choice but a type
of the lower-middle of Fig.~\ref{fig:conventional}.

\begin{figure}
 \begin{center}
 \includegraphics[width=6cm]{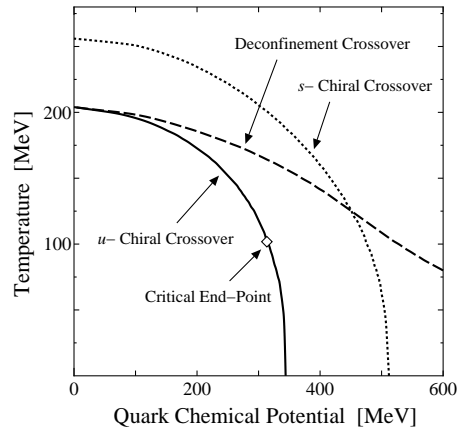}
 \end{center}
\caption{Phase diagram in the three-flavor PNJL model taken from
  Ref.~\cite{Fukushima:2008wg}.}
\label{fig:phase}
\end{figure}

  Indeed, the phase diagram in the three-flavor PNJL model takes an
uncommon structure as given in Fig.~\ref{fig:phase}.  The boundary
lines are determined by the condition that
$\langle\bar{u}u\rangle/\langle\bar{u}u\rangle_0=0.5$ (solid line) and
$\ell=0.5$ (dashed line) with $\langle\bar{u}u\rangle_0$ being the
chiral condensate in the vacuum.  The dotted line represents the
chiral crossover associated with strange quarks which spreads over
higher temperature due to strange quark mass.

  It is remarkable that the locking of deconfinement and chiral
restoration no longer persists in the high-$\mu$ region.  The physical
meaning of the region where $\langle\bar{u}u\rangle\neq0$ and
$\ell\simeq0$ is not quite straightforward;  the system is nearly
center symmetric but the relevant degrees of freedom are presumably
quarks.  Such weird quark matter is named the ``quarkyonic phase'' by
McLerran and Pisarski~\cite{McLerran:2007qj} in the context of the
large $N_c$ limit.  (See also discussions in
Ref.~\cite{Glozman:2007tv}.)


\section{To what extent can we believe model predictions?}

  In the three-flavor PNJL model the QCD critical point is located at
$T=102\;\mbox{MeV}$ and $\mu=313\;\mbox{MeV}$.  I have to emphasize
that this model output (or any other model outputs as well) is nothing
more than one suggestive indication and should not be regarded as
conclusive.

  The low-energy effective model cannot avoid ambiguity involving the
model parameter.  In a usual prescription the model parameter is
fixed at $T=\mu=0$ and assumed to be constant at any $T$ and $\mu$.
Hence the thermal excitation of the effective degrees of freedom alone
yields the $T$ and $\mu$ dependent contributions.  This assumption is
maybe acceptable as long as $T$ and $\mu$ are smaller than the model
cutoff scale $\Lambda$ which characterizes the relevant energy scale
of the effectiveness.  This is because the model parameter encompasses
the microscopic dynamics of more elementary particles and those
particles reside in the energy scale higher than $\Lambda$, so that
their excitation effect should be negligible at $T$ and $\mu$ smaller
than $\Lambda$.

  There are, however, some model parameters that would significantly
change the resultant physical predictions.  Here I shall briefly
address two such examples according to Ref.~\cite{Fukushima:2008wg}.

  The first example is the $\mathrm{U_A}(1)$ symmetry breaking term
which is often called the 't~Hooft interaction.  This term has an
origin in the instanton distribution.  Because the finite $T$ and
$\mu$ effects suppress the instanton excitation, the 't~Hooft
interaction should diminish exponentially.

\begin{figure}
 \includegraphics[width=6cm]{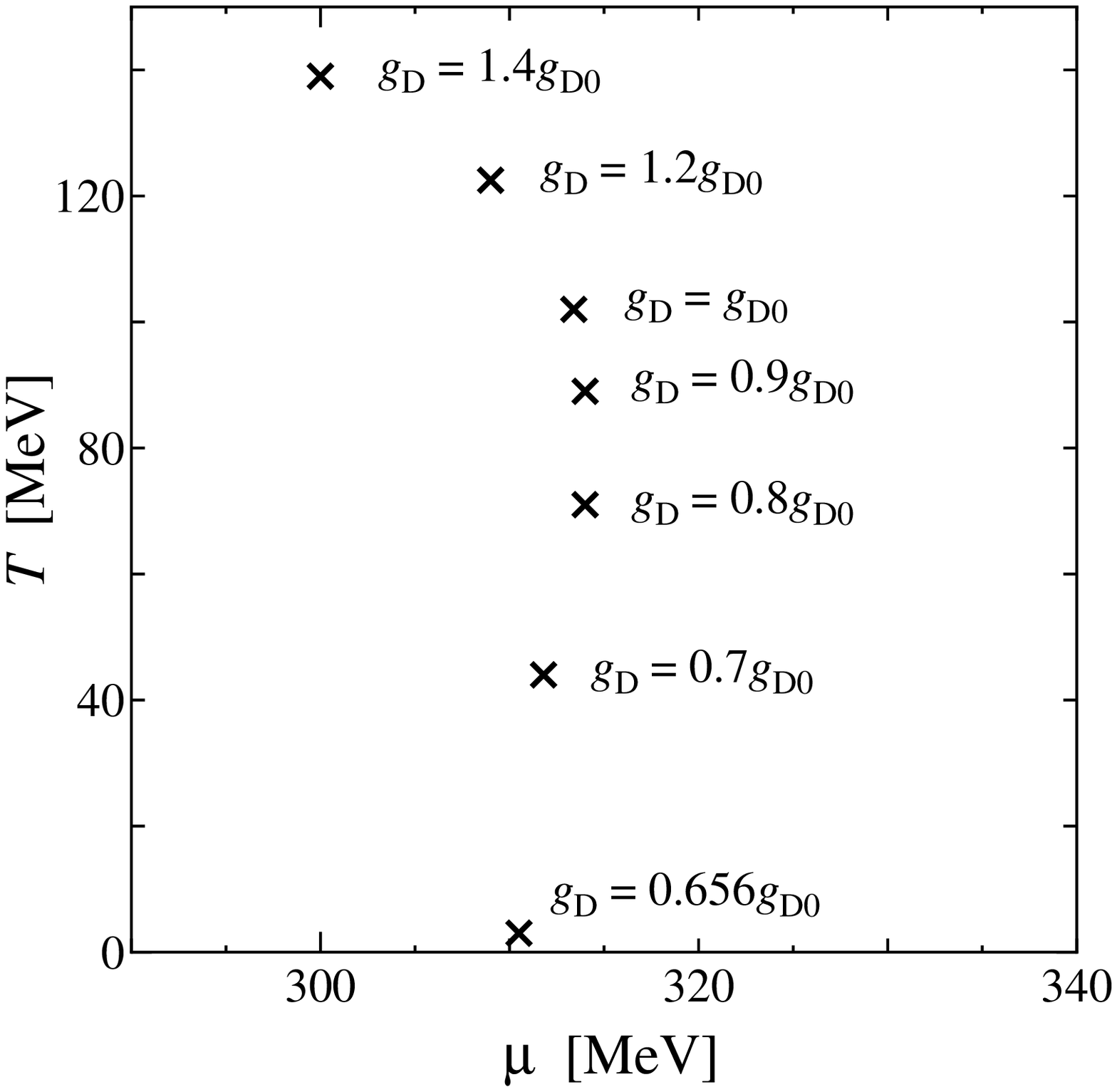}
 \hspace{5mm}
 \includegraphics[width=5.8cm]{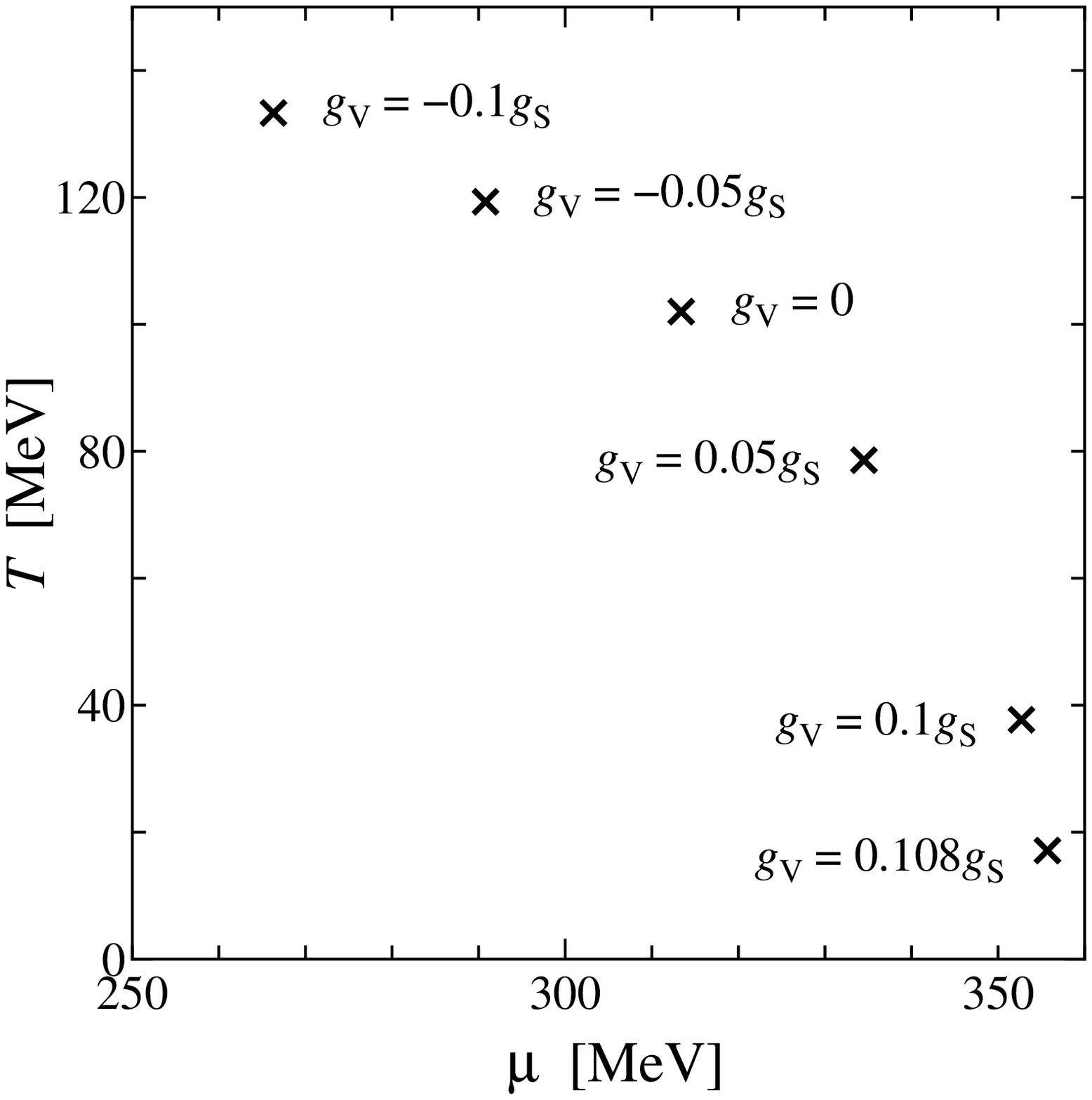}
 \caption{QCD critical point in the PNJL model as a function of the
 uncontrolled model parameters.  Left) 't~Hooft interaction
 dependence.  Right) Vector-channel interaction dependence taken from
 Ref.~\cite{Fukushima:2008wg}.}
 \label{fig:nocep}
\end{figure}

  The strength of the first-order phase transition strongly depends on
the magnitude of the $\mathrm{U_A}(1)$ symmetry breaking in the case
of three-flavor quark matter.  Thus, if it gets smaller by the
finite-$\mu$ effect, the first-order phase transition region should
shrink, and the QCD critical point should move closer to the $T=0$
axis.  It will eventually disappear from the phase diagram with the
't~Hooft interaction below a certain critical value.  The left of
Fig.~\ref{fig:nocep} shows the location of the QCD critical point as a
function of the $\mathrm{U_A}(1)$ symmetry breaking strength denoted
by $g_{\rm D}$ (and $g_{\rm D0}$ being the vacuum value thereof).

  It is surprising to see that only $35\%$ reduction of $g_{\rm D}$
can be enough for the QCD critical point to disappear.  There is no
quantitative estimate of how much $g_{\rm D}$ should be reduced at
finite $T$ and $\mu$ (see Ref.~\cite{Fukushima:2001hr} for an
attempt), but $35\%$ is within a possible reach as compared to the
exponential suppression.

  The second example is the strength of the vector-channel
interaction.  The right of Fig.~\ref{fig:nocep} indicates the
sensitivity of the QCD critical point as a function of the
vector-channel interaction denoted by $g_{\rm V}$.  Since the zeroth
component in the vector channel directly couples the density, the
vector-channel interaction could be induced by the finite-$\mu$
environment, which is not quite well known from theory.  In the plot
of the right of Fig.~\ref{fig:nocep} the interaction strength is given
in the unit of the scalar-channel interaction $g_{\rm S}$.  Taking
account of the Fierz transformation, $g_{\rm V}$ could be as large as
$g_{\rm S}$.  The critical value, $g_{\rm V}\simeq0.11g_{\rm S}$, is
surprisingly small in this sense.

  In view of these above results our conclusion should be that the
location of the QCD critical point and even its existence are not
under theoretical control and thus not robust.  In fact, a recent
lattice simulation goes negative against the existence of the QCD
critical point~\cite{forcrand}.


\section{Remarks}

  Finally let me comment on the aspects of chiral symmetry and its
dynamics relevant to the heavy-ion collisions.

  In general the chiral symmetry breaking and restoration would have
an effect mostly on the hadron mass spectrum.  The bulk property of
hot and dense matter is thus insensitive to chiral dynamics.
Consequently the chiral phase transition, if any, cannot bring about a
perceivable change in the hydrodynamic evolution.  One exception might
be the QCD critical point where the equation of state has
singularity leading to the focusing effect~\cite{nonaka}.

  Hence, it should be not the bulk thermodynamics but the collective
excitations that are more pertinent quantities to probe the chiral
properties.  The dilepton spectrum and the associated mass shift
and/or broadening would be a nice candidate to disclose the chiral
characteristics.

  One might then think that the dilepton measurement could pinpoint
the QCD critical point by the presence of the massless $\sigma$ meson.
Unfortunately this idea would not work out because the criticality
lies in the space-like channel, while the decay channel is time-like.
The spectral function in the space-like region may behave
discontinuously from the time-like region~\cite{fujii}.  Therefore, so
far, a promising signature for the QCD critical point is the
fluctuation associated with the soft mode in the vicinity of the
critical point (see also Ref.~\cite{Asakawa:2008ti} for a recent
proposal for another signature).  As a final remark, I would like to
repeat that we should always keep in our mind that the QCD critical
point is still a hypothetical idea and its location and existence
should be tested by experiments.

  This work is in part supported by Yukawa International Program for
Quark-Hardon Sciences.

\end{document}